\documentclass{article}
\usepackage{dcase2024}
\usepackage{amsmath,graphicx,url,times,booktabs, tabularx}
\usepackage{gensymb}
\usepackage{mdframed}
\usepackage{ragged2e} 
\usepackage{xcolor}
\usepackage{makecell}

\title{Towards Spatial Audio Understanding via Question Answering}

\name{Parthasaarathy Sudarsanam,
      Archontis Politis
      }
\address{Audio Research Group, Tampere University, Tampere, Finland}

\begin{document}

\ninept
\maketitle

\begin{sloppy}

\begin{abstract}
In this paper, we introduce a novel framework for spatial audio understanding of first-order ambisonic (FOA) signals through a question answering (QA) paradigm, aiming to extend the scope of sound event localization and detection (SELD) towards spatial scene understanding and reasoning. First, we curate and release fine-grained spatio-temporal textual descriptions for the STARSS23 dataset using a rule-based approach, and further enhance linguistic diversity using large language model (LLM)-based rephrasing. We also introduce a QA dataset aligned with the STARSS23 scenes, covering various aspects such as event presence, localization, spatial, and temporal relationships. To increase language variety, we again leverage LLMs to generate multiple rephrasings per question. Finally, we develop a baseline spatial audio QA model that takes FOA signals and natural language questions as input and provides answers regarding various occurrences, temporal, and spatial relationships of sound events in the scene formulated as a classification task. Despite being trained solely with scene-level question answering supervision, our model achieves performance that is comparable to a fully supervised sound event localization and detection model trained with frame-level spatiotemporal annotations.
The results highlight the potential of language-guided approaches for spatial audio understanding and open new directions for integrating linguistic supervision into spatial scene analysis.

\end{abstract}


\section{Introduction}
\label{sec:intro}


Scene understanding by machines using audio signals is a well-established problem in audio processing, audio-based machine learning, and the broader field of machine learning for natural scenes. Early and widely studied tasks include the classification of scene types from audio recordings and the detection of sound events over time corresponding to specific target classes \cite{mesaros2021sound}. While these approaches provide valuable semantic insights into the content of an auditory scene, they lack information about the spatial characteristics of the sound environment. To address this limitation, the task of sound event localization and detection (SELD) was introduced \cite{adavanne2018sound, politis2020overview}. SELD extends traditional methods by capturing both the temporal activity of sound events and their spatial locations relative to the recording device. This spatial information is essential for downstream applications that depend on the positioning and spatial relationships of sound sources within a scene. Introduced as part of the DCASE Challenge in 2019, SELD has built an active and growing research community dedicated to advancing SELD methods. \cite{nguyen2021general, shimada2021accdoa, wang2023four, hu2024selective, shul2024cst}.

Most SELD proposals that are capable of handling dynamic complex scenes follow a strongly supervised training paradigm where event activity and location labels are provided at a fine temporal resolution, with a few exceptions trying to leverage self-supervision \cite{santos2024w2v, jiang2024exploring}. Similarly, during inference SELD models are expected to 
provide predictions at a similar resolution. Such annotations are very difficult to obtain in real scenes, with only a handful of such datasets currently existing \cite{brousmiche2020secl, politis2022starss22}, including the STARSS22-23 dataset collected by the authors and collaborators \cite{politis2022starss22,shimada2023starss23}. Otherwise, supervised training of SELD methods has relied on simulations of spatial sound scenes \cite{politis2021dataset, roman2024spatial}.


A recent trend in scene understanding across domains involves grounding perception in natural language \cite{CLIP_Radford2021, CLAP_Elizalde2023}. It has been explored in a few works for spatial audio scene understanding \cite{zheng2024bat, devnani2024learning, tang2024can, xie2025thinking}. Focusing on general sound events, the BAT system   \cite{zheng2024bat} evaluates a model’s ability to answer questions on classification, detection, direction, and distance using simulated binaural recordings with up to two static events in reverberant scenes. Questions and answers were generated based on a rule-based approach, while an LLM was used for the QA task. The ELSA system \cite{devnani2024learning} trains a spatial audio model using contrastive learning on a large simulated dataset of FOA recordings paired with captions describing spatial properties. Due to the lack of captioned spatial audio data, the authors use standard audio captioning datasets, simulate FOA by placing sounds in rooms, and revise the original captions with a language model to include source position and room size.


In this work, we train a model that answers natural language questions about sound event localization and detection information, including sound event presence, classification, and temporal and spatial order, based on the sound scene. Unlike previous studies, our model is trained and evaluated on FOA audio from real scene recordings. We use the STARSS23 dataset, which contains approximately eight hours of audio with fine grained spatiotemporal annotations at 100ms intervals. Firstly, we generate a set of detailed spatial captions by converting the metadata into textual descriptions that capture the evolving spatiotemporal structure of each scene, and enhance them using GPT-4 for linguistic diversity. These captions can support both spatial audio question answering and general scene understanding tasks. Secondly, we construct a QA dataset for STARSS23 and similarly apply GPT-4 to rephrase rule-based questions. Finally, we develop a baseline spatial audio question answering model that takes FOA audio features and natural language questions as input, and frames the task as classification.

\begin{table*}[t]
\centering
\caption{Types of questions in the STARSS23 QA dataset}
\label{tab:qa_types}

\begin{tabular}{c|l|l}
\toprule
\textbf{Type} & \textbf{Question Type} & \textbf{Description} \\
\midrule
I&Yes/No Detection & Check presence of a specific sound event. \\
II&Event Listing & List all sound events in the clip.  \\
III&Spatial Understanding & List all stationary or moving sound sources.  \\
& & Find the leftmost, rightmost, topmost, bottommost, nearest, or farthest sound event. \\
IV&Spatial Relationship & Sort events by azimuth, elevation, or distance in ascending or descending order. \\
V&Temporal Relationship & Sort events based on order of appearance. \\
\bottomrule
\end{tabular}
\end{table*}

\section{Dataset creation}
\subsection{STARSS23 Captions dataset}
\label{sec:format}

To generate detailed textual descriptions of the recorded scenes, we utilize the annotations provided in the STARSS23 dataset. These annotations are available at a temporal resolution of 100 ms and consist of the following fields: \texttt{frame time}, \texttt{sound event label}, \texttt{parent source ID}, \texttt{azimuth angle}, \texttt{elevation angle}, and \texttt{distance}. Annotations reflect multiple sound events occurring at the same time, including events of the same class belonging to different sources.

The durations of the individual recordings in STARSS23 vary between 30 seconds to 9 minutes. To standardize processing, we first segment each recording into 10-second clips. We then apply a rule-based algorithm to convert the per-frame annotations into structured textual descriptions as follows: \vspace{0.1cm}

    \noindent\textbf{Frame-Level Descriptions:} For each annotated frame, we generate a basic description of the form: \textit{“From \texttt{start\_time} to \texttt{end\_time}, a \texttt{sound\_event\_label} is heard. Horizontal angle \texttt{azimuth}, vertical angle \texttt{elevation}, distance \texttt{distance}, source ID: \texttt{parent\_source\_id}.”}
    
    \noindent\textbf{Sorting by Source:} The frame-level descriptions are grouped and sorted based on the \texttt{parent\_source\_id}. \vspace{0.1cm}
    
    \noindent\textbf{Sorting by Sound Event:} Within each source group, descriptions are further sorted by \texttt{sound\_event\_label}. \vspace{0.1cm}
    
    \noindent\textbf{Instance Segmentation:} Contiguous frame descriptions (with 100 ms intervals) are grouped into instances. Any break in this interval indicates a new instance, allowing segmentation of repeated but temporally separated sound events from the same source.\vspace{0.1cm}
    
    \noindent\textbf{Trajectory-Based Description:} For each instance, we construct a textual summary that captures both temporal and spatial dynamics. Specifically, we log the instance's onset and offset times, along with the starting and ending values of azimuth, elevation, and distance. Additionally, we track the minimum and maximum values of these spatial parameters and the corresponding timestamps at which they occur. This enables us to generate an approximate trajectory-based description of the sound source's motion over time. For static events these values are all equal. \vspace{0.1cm}

The following is a template-based generated description for one instance of a sound event based on the annotations:  \textit{From 0.2s to 1.4s, man speaking is heard. It is initially at an azimuth angle of -70 degrees and moved finally to an azimuth of -95 degrees. During this time, the sound source moved to a maximum azimuth angle of -70 degrees at 0.3s and to a minimum azimuth angle of -95 degrees at 1.3s. The sound was coming throughout from an elevation angle of approximately -46 degrees. The sound was coming throughout from a distance of approximately 97cm. Source id: 2}

Since this rule-based approach results in limited linguistic diversity, we use GPT-4 to paraphrase the generated captions, thereby producing a more diverse set of textual descriptions. The Prompt used to generate the GPT captions is as follows:


\noindent \textit{Paraphrase the following sentences while preserving all factual details, numbers, and structure of the events. Do not add any new information or descriptions. Ensure the paraphrase sounds natural and fluent, but all temporal, spatial, and entity-related details (time, azimuth, elevation, distance, source ID) must remain exactly the same.}\newline
\noindent\textit{ Sentence:} \\
\noindent\textit{\textit{ Template-based textual description}} \vspace{0.1cm}

As an example, GPT-4 produces the following textual description from the earlier rule-based instance example: \textit{Between 0.2s and 1.4s, the sound of a man speaking is heard. The source starts at an azimuth angle of -70 degrees and ends at -95 degrees. During this interval, it reaches its highest azimuth of -70 degrees at 0.3s and its lowest of -95 degrees at 1.3s. The elevation angle remains steady at approximately -46 degrees throughout. The sound is continuously perceived from a distance of around 97cm. Source ID: 2}

\subsection{STARSS23 QA dataset}
\label{sec:pagelimit}

We created a question-answering (QA) dataset for the 10-second segments of the STARSS23 data. Rule-based scene descriptions were used to generate corresponding answers. The QA dataset includes several types of questions, summarized in Table~\ref{tab:qa_types}. 

Each question type is designed to assess a specific aspect of acoustic scene understanding, from typical sound event classification tasks to more complex spatial and temporal reasoning. For instance, Yes/No detection questions (Type I) determine whether a particular sound event occurs within a clip, while event listing questions (Type II) aim to identify all detected sound events. Spatial understanding questions (Type III) focus on identifying absolute spatial properties of sound sources, such as determining which source is closest or farthest, lowest or highest, or leftmost or rightmost in the scene, or listing all moving or stationary events in the scene. Spatial Relationship questions (Type IV) require the model to understand the sound events comparatively, asking it to order all sound sources based on various spatial attributes. Similarly, Temporal Relationship questions (Type V) assess the model’s ability to understand the temporal sequence of sound events by asking it to order all sound events based on when they occur. 

Table~\ref{tab:qa_examples} shows example questions and answers for each of the question types described above. These examples demonstrate the diversity and depth of the QA dataset, ranging from simple detection and listing tasks to more sophisticated reasoning about spatial layouts and temporal orderings.  While it is possible to formulate many additional question types around spatial and temporal relationships, we limit this exploratory work to the types outlined above. For each question, we also generated 10 linguistically diverse variations using GPT-4 to promote variability and robustness in language understanding.

\begin{table*}[h]
\centering
\caption{Example questions and answers from the STARSS23 QA dataset}
\label{tab:qa_examples}

\begin{tabular}{c|l|l}
\toprule
\textbf{Type} & \textbf{Example Question} & \textbf{Example Answer} \\
\midrule
I & Is there a sound event of a telephone ringing in the scene?
 & Yes \\
II & Which sound sources are active? & ['woman speaking', 'man speaking'] \\

III & Which sound event is the farthest from the microphone? & 'laughing' \\
      & What sound sources remain stationary in this scene? & ['music', 'bell']  \\
IV & Order the audio events by distance, beginning with the closest.
 & ['footsteps', 'music', 'door open or close'] \\
   & Sort the audio events from the bottommost to the topmost.
 & ['footsteps', 'clapping', 'man speaking] \\
V & Arrange the sound sources in order of when they begin, from earliest to latest. & ['door open or close', 'footsteps', 'man speaking'] \\
\bottomrule
\end{tabular}
\end{table*}

Since STARSS23 consists of real-world recordings, multiple instances of the same sound event from the same source can occur within a single clip. Additionally, there are cases where moving sound sources cross paths in space. To simplify the QA task, we only consider the first appearance of each source in the scene when constructing the answers, using the initial time and position information for any required sorting or temporal reasoning.

\section{Methods}
\label{sec:pagestyle}

\subsection{SELD QA Model}

We formulate the QA task as a classification problem and train an encoder-decoder model to address it. The encoder is based on the SELDnet architecture \cite{adavanne2018sound}, which includes convolutional and recurrent layers followed by a self-attention module. These layers extract time-frequency features from multichannel input audio while preserving spatial information. The self-attention block models longer-term dependencies in the scene and refines the representation of spatial cues. The output of this encoder captures the relevant acoustic and spatial context of the scene and is passed to the decoder.

The decoder is implemented as a six-layer transformer model with an attention size $d\_model = 512$ trained from scratch. It takes as input the encoded representation of the audio scene along with a natural language question. The question is first mapped to a continuous vector space using pre-trained FastText word embeddings~\cite{mikolov2018}, each with 300 dimensions. These embeddings help capture the semantics of different phrasings of a question. The transformer decoder attends jointly to the audio features and the embedded question to produce the final prediction.

The task is formulated as a multi-class classification problem. For the yes or no questions, the output layer contains two neurons. For event presence, spatial ordering, and temporal ordering tasks, the decoder includes \(N\) neurons for each task, where \(N\) is the total number of unique sound event classes. As a result, the output layer consists of \(3N + 2\) neurons, enabling the model to handle a variety of question types. Figure \ref{fig:seld_qa_architecture} shows the model architecture of the SELD QA model.

\begin{figure}
    \centering
    \includegraphics[width=\linewidth]{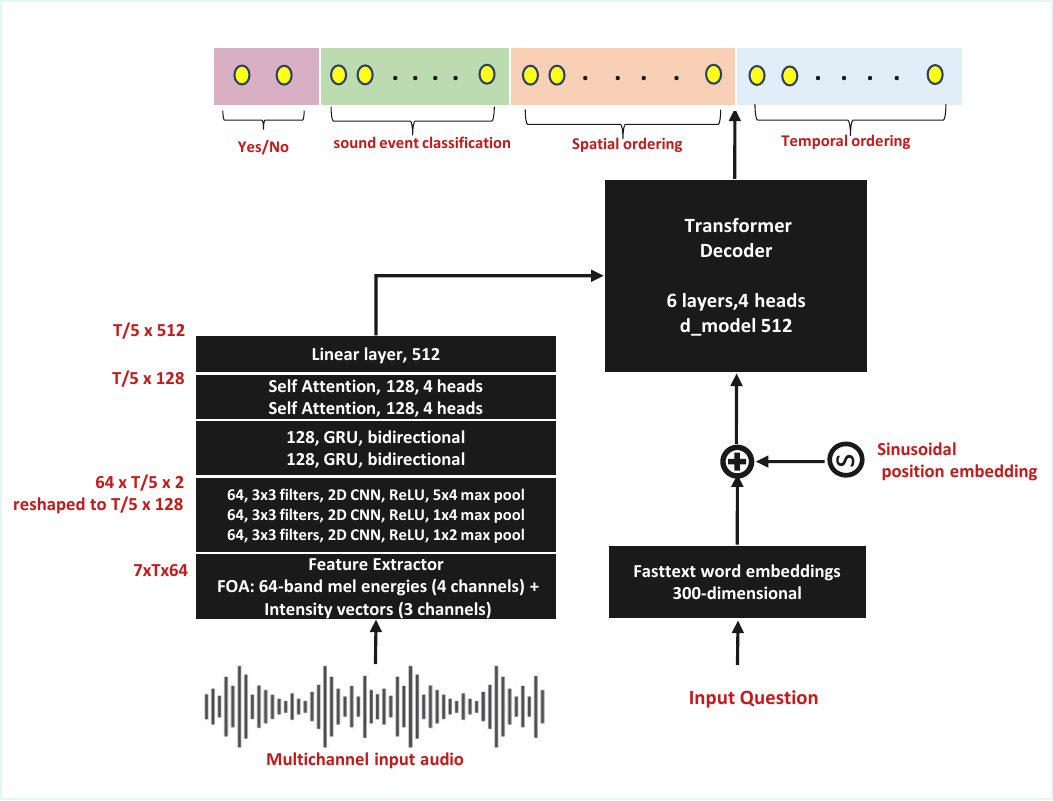}
    \caption{The proposed SELD QA model architecture}
    \label{fig:seld_qa_architecture}
\end{figure}

\subsection{Loss function}

The proposed SELD-QA model jointly handles sound event classification as well as spatial and temporal ordering tasks. Accordingly, we design a loss function composed of binary cross-entropy loss for sound event classification and ranking-based losses for ordering predictions.

For sound event classification questions, the binary cross-entropy (BCE) loss $\mathcal{L}_{\text{BCE}}$ is applied to the output neurons corresponding to each sound event class, as well as to the dedicated yes and no neurons. This loss is computed using the binary ground truth, indicating whether each sound event is active in response to the given question. Note that the BCE loss is also computed for ordering-based questions to ensure that the model accurately detects the relevant sound events while attempting to predict their spatial or temporal order.

For spatial and temporal ordering tasks, we use a combination of a pairwise ranking loss and an L1 regression loss. These are applied separately to spatial ordering neurons and temporal ordering neurons depending on the question type


\begin{align}
 \mathcal{L}_{\text{spatial/temporal}} &= \mathcal{L}_{\text{rank}}^{\text{spatial/temporal}} + \mathcal{L}_{\text{L1}}^{\text{spatial/temporal}}
\end{align}
 
The total loss used to train the model is the sum of the sound event classification loss and the ordering losses:

\begin{equation}
 \mathcal{L}_{\text{total}} = \mathcal{L}_{\text{BCE}} + \mathcal{L}_{\text{spatial}} + \mathcal{L}_{\text{temporal}}
 \end{equation}
 The goal of the ranking and L1 losses is to ensure that sound events predicted as active are ranked higher than inactive ones, and, within the set of active events, their predicted ranks match the ground-truth temporal or spatial order. For the $i$-th temporal or spatial ordering output, the model provides an ordering score $p_i\in[0, 1]$, where a score of 0 indicates that no sound event occurs for the $i$-th class while higher values indicate a higher position in the relative spatial or temporal ordering compared to another sound event with a lower non-zero score. The ground truth ordering is also encoded in the same way.

Regarding the ranking loss, we define $\mathcal{A}=\{a_1,a_2,..,a_M\}$ as the ordered list of $M$ active sound classes representing the correct temporal (e.g. from earliest to latest) or spatial (e.g. from left to right) order, with $a_m\in{1,..,N}$ being the class index. We additionally define the set $\mathcal{B}$ containing the indices of the $N-M$ inactive sound classes. The ranking loss is then defined as




\begin{equation}
\begin{split}
\mathcal{L}_{\text{rank}} = 
& \sum_{i=1}^{M} \sum_{j \in \mathcal{B}} \max\big(0,\ \delta - (p_{a_i} - p_j)\big) \\
& + \sum_{i=1}^{M-1} \sum_{j=i+1}^{M} \max\big(0,\ \delta - (p_{a_i} - p_{a_j})\big)
\end{split}
\end{equation}

\noindent where \( \delta \) is a fixed margin, set to 0.3 in our case. The first term promotes higher scores for active over inactive classes, while the second enforces correct ordering among the active classes. Additionally, the L1 loss is used to regress toward the ideal ordering values. 
 
 \begin{equation}
 \mathcal{L}_{\text{L1}} = \sum_{i=1}^N|\hat{p}_i - p_i|.
 \end{equation}

\noindent where $N$ is the total number of classes and $\hat{p}_i$ is the ground truth ordering score of the class $i$. These components ensure that the model learns to detect sound events accurately while also reasoning about their relative spatial and temporal structure when required.

\section{Evaluation}

We train our model using the STARSS23 dev-train split and the dev-train-synth dataset provided in the DCASE2024 Challenge. For reproducibility, the evaluation is conducted on the publicly available STARSS23 dev-test split. Our model is trained for $100$ epochs using the AdamW optimizer with a learning rate of $1e-5$. For a fair comparison, we evaluated our model against the DCASE2024 audio-only SELD baseline model, which is trained on the same dataset without any data augmentation. Note that this baseline is trained using 100ms frame wise labels for SELD task, following a regression setup with the Multi-ACCDOA loss \cite{shimada2022multi} to estimate DoA and distances. The baseline model's predictions are then post-processed using the same rule-based approach used in the QA dataset construction to derive answers for the QA tasks.

To evaluate the performance of our model, we use different metrics tailored to the specific question types. For sound event detection questions (Type I-III), we report the standard precision, recall, and F1 score to assess the model's ability to accurately identify active sound classes. 

For spatial and temporal ordering questions, we adopt a modified version of the Mean Reciprocal Rank \(\text{MRR}_{\text{mod}}\), which compares the predicted order of sound event classes (filtered to include only predicted active classes) with the ground-truth order. For each active target class, we compute the positional difference between its predicted and actual rank, assigning higher rewards to more accurate placements. The \(\text{MRR}_{\text{mod}}\) is given by

\begin{equation}
\text{MRR}_{\text{mod}} = \frac{1}{n} \sum_{c \in T}
\begin{cases}
\frac{1}{1 + |\hat{r}_c - r_c|}, & \text{if } c \in \hat{T} \\
0, & \text{if } c \notin \hat{T}
\end{cases}
\end{equation}

\noindent Here, \( T = [c_1, c_2, \dots, c_n] \) denotes the ordered list of \( n \) unique ground truth active sound event classes corresponding to a given question. The set \( \hat{T} \) denotes the predicted ordering of sound event classes identified as active by the model. For each class \( c \in T \), \( r_c \) specifies its rank in the ground truth ordering, and \( \hat{r}_c \) denotes its predicted rank in \( \hat{T} \). If a class \( c \) is not present in \( \hat{T} \), it receives a score of zero. Predictions in the ordering output that include classes not active in the scene are implicitly penalized through the precision, recall, and F1 metrics.

\section{Results}

Table~\ref{tab:results} presents the performance comparison between the DCASE2024 baseline SELD model and the proposed QA-based SELD model. The baseline model is trained using framewise supervision for sound event detection and localization, while our model is trained using question-answer pairs derived from the same annotations.

\begin{table}[!h]
    \centering
    \small
    \resizebox{\linewidth}{!}{
    \begin{tabular}{c|c|c|c|c|c}
    \toprule
        \textbf{Model} & \textbf{Precision} & \textbf{Recall} & \textbf{F1} & \textbf{\makecell{Spatial\\$\mathrm{MRR}_{\mathrm{mod}}$}} & \textbf{\makecell{Temporal\\$\mathrm{MRR}_{\mathrm{mod}}$}} \\
        \midrule
        Baseline SELD & 0.74 & 0.76 & 0.74 & 0.58 & 0.59 \\
        SELD QA    & 0.73 & 0.69 & 0.70 & 0.56 & 0.59 \\
        \bottomrule
    \end{tabular}}
    \caption{Performance comparison between the DCASE2024 baseline SELD model trained with framewise label supervision and the proposed SELD-QA model.}
    \label{tab:results}
\end{table}

It can be seen that the proposed SELD QA model achieves comparable performance to the baseline SELD model across all evaluation metrics. While the baseline achieves a slightly higher F1 score of 0.74 compared to 0.70, and marginally better spatial MRR\textsubscript{mod}, both models perform identically in terms of temporal MRR\textsubscript{mod}.

It is important to note that the baseline model is trained using fine-grained, frame-wise labels for both sound event detection and localization, providing dense supervision throughout the audio sequence. In contrast, our QA-based SELD model is trained using scene-level question-answer pairs, which provide only coarse, overall supervision without requiring temporally aligned annotations. Despite this weaker supervision, our model maintains competitive performance, demonstrating the effectiveness of the QA-based formulation for SELD tasks and its potential for reducing annotation effort in real-world scenarios.

\section{Conclusion}

In this work, we introduced a spatiotemporal captioning dataset and a corresponding QA dataset built from the STARSS23 real scene recordings. These resources support a novel formulation of the SELD task as a QA problem that probes event detection as well as the relative spatial and temporal relationships of sound sources within a scene. This approach opens new directions for evaluating machine understanding of complex acoustic environments. We developed a baseline QA model that frames spatial audio question answering as a classification task. Our results show that this model performs comparably to the DCASE SELD baseline model trained with frame-level supervision. For future work, we aim to move beyond classification by generating descriptive natural language answers. We also plan to explore pretrained spatial audio encoders and large language model decoders to improve performance on spatial question answering and advance text grounded spatial audio understanding.

\section{Acknowledgement}

The funding for this work is supported by Jane and Aatos Erkko Foundation through the CONVERGENCE of Humans and Machines project. This work was also supported in part by a grant from the Finnish Foundation for Technology Promotion.
\bibliographystyle{IEEEtran}
\bibliography{refs_AP}

%
%
%
%
%
%
%
%
%

\end{sloppy}
\end{document}